\documentclass{article}

\usepackage[preprint,nonatbib]{neurips_2022}

\usepackage[utf8]{inputenc} 
\usepackage[T1]{fontenc}    
\usepackage{hyperref}       
\usepackage{url}            
\usepackage{booktabs}       
\usepackage{amsfonts}       
\usepackage{nicefrac}       
\usepackage{microtype}      
\usepackage{xcolor}         
\usepackage[style=numeric,sorting=none]{biblatex}  
\usepackage{adjustbox}
\usepackage{amsmath}
\usepackage{multirow}
\usepackage{color}
\usepackage{algorithm}
\usepackage{algorithmic}
\usepackage{setspace}
\usepackage{wrapfig}
\usepackage{enumitem}
\usepackage{caption}
\usepackage{subfigure}
\setlist{topsep=0pt, leftmargin=*}

\title{Do Deep Learning Methods Really Perform Better in Molecular Conformation Generation?}

\author{%
Gengmo Zhou$^{1,2}$\thanks{\scriptsize Equal contribution.}~, Zhifeng Gao$^{1}$\footnotemark[1]~, Zhewei Wei$^{2}$, Hang Zheng$^{1}$, Guolin Ke$^{1}$\thanks{\scriptsize Corresponding authors.} \\
$^{1}$DP Technology, China\\
$^{2}$Renmin University of China, China\\
\texttt{\{zhougm, gaozf\}@dp.tech} \\
\texttt{zhewei@ruc.edu.cn} \\
\texttt{\{zhengh, kegl\}@dp.tech} \\
}

\begin{document}

\maketitle
\begin{abstract}
Molecular conformation generation (MCG) is a significant and vital problem in drug development. While traditional methods, like systematic searching, model-building, random searching, distance geometry, molecular dynamics, and Monte Carlo methods, have been developed to solve the MCG problem, they have limitations that are dependent on the molecular structures. Recently, deep learning-based MCG methods have been introduced, claiming to outperform the traditional methods. Surprisingly, we have developed a simple and cheap algorithm (parameter-free) based on traditional methods and found that it is comparable to, or even outperforms, deep learning-based MCG methods on widely used GEOM-QM9 and GEOM-Drugs benchmarks. Our algorithm design involves clustering of the RDKit-generated conformations. We believe that our findings can help the scientific community to revise deep learning methods and benchmarks for MCG. The code for the proposed algorithm is available at \url{https://gist.github.com/ZhouGengmo/5b565f51adafcd911c0bc115b2ef027c}.
\end{abstract}

\section{Introduction}
\label{intro}

Molecular conformation generation (MCG) plays a crucial role in drug discovery~\cite{crum1865connection,hansch1964p}, as it aims to generate low-energy spatial arrangements of atoms in a molecule. MCG is closely related to various drug design tasks, including molecular property prediction~\cite{axelrod2022geom, fang2022geometry}, protein-ligand complex~\cite{roy2015understanding, morris2009autodock4, trott2010autodock}, structure similarity~\cite{hawkins2007comparison, roy2015ligsift}, and pharmacophore searching~\cite{schwab2010conformations}.

The traditional approach for MCG~\cite{hawkins2017conformation} involves using conformational search methods to generate a set of molecular conformations, followed by applying energy minimization methods to optimize them further using energy estimators, such as molecular mechanics (MM) or density functional theory (DFT)~\cite{parr1980density, baseden2014introduction}. Finally, a batch of optimized conformations with low energies are screened out based on their potential energy. Various traditional approaches have been proposed for different scenarios.

For instance, RDKit, a popular cheminformatics software in the chemical community, is widely used in MCG. RDKit uses ETKDG~\cite{riniker2015better}, a template-based method, in its default setting to generate conformations efficiently. Then, MM methods such as MMFF~\cite{halgren1996merck} or UFF~\cite{rappe1992uff} are applied to perform structure optimization and obtain better conformations. Although a typical and efficient MCG method, it suffers from low performance in sampling coverage due to the limitations of template-based sampling and in energy evaluation due to the low accuracy of MM force fields.

Recently, several works applied deep learning to MCG, aiming to generate widely distributed and low energy conformations efficiently. These works can be classified into two categories. One category is methods based on distance geometry~\cite{simm2020generative, xu2020learning, xu2021end, shi2021learning, luo2021predicting}, which first generates atoms' pair-distance matrix and then recoveries atoms' coordinates based on the matrix. Another category is methods that generate atoms' coordinates directly~\cite{mansimov2019molecular,ganea2021geomol,zhu2022direct,xu2022geodiff}. With the advent of diffusion models, there are also diffusion based methods~\cite{xu2022geodiff} proposed for MCG tasks.

In these deep learning models, GEOM\cite{axelrod2022geom} dataset is widely used for training and evaluation. It uses extensive sampling methods, based on the semi-empirical DFT, to generate 3D conformations. In practice\cite{shi2021learning}, 40k molecules (200k conformations) are sampled from GEOM for training, and 200 molecules (about 20k conformations) are sampled for testing.
As there are multiple conformations for one molecule, Coverage (COV) and Matching (MAT) are proposed to evaluate the performance of these models. In particular, COV is the coverage of the generated conformation to the reference one under a certain RMSD threshold; the larger, the better diversity. MAT measures the difference between the generated conformation and the reference one; the smaller, the better accuracy.

The deep learning models demonstrated superior performance compared to traditional methods in the aforementioned benchmark. However, despite the models being proposed several years ago, we could not identify any real-world applications utilizing these deep models for conformation generation. Consequently, we question whether these models truly surpass traditional approaches. 

Remarkably, we develop a straightforward algorithm based on traditional techniques that outperforms most deep learning models. Specifically, we employed the EDKTG algorithm integrated in RDKit \cite{landrum2013rdkit} and utilized three samplers: uniform, geometric, and energy samplers to generate diverse and low-energy conformations. We then applied an unsupervised clustering algorithm to obtain representative conformations. 

Our unexpected discovery challenges the superiority of existing deep learning models. As a result, we reassess these methods and offer the following suggestions for future research. 

\begin{itemize}
    \item \emph{The current benchmark may be wrong.} Our simple algorithm outperformed all previous deep models with ease, indicating that the current benchmark may be vulnerable to manipulation. We suspect that a more diverse range of conformations could achieve even better results, and have designed an additional experiment to explore this possibility explicitly. 
    \item \emph{The benchmark should focus on the end applications, not the standalone conformations.} Molecular conformations are highly flexible, varying significantly depending on the environment, and can be difficult to obtain through wet experiments. Additionally, different applications have different requirements for conformations. For example, in molecular docking \cite{roy2015understanding, morris2009autodock4,trott2010autodock}, the goal is to find conformations that are close to the binding conformation of the target, while for quantum property prediction, the goal is to identify conformations with the lowest energy. Therefore, benchmarking the effect of conformation generation should be based on the specific requirements of downstream applications, or on the final performance of these applications.

\end{itemize}
\section{Related work}

For MCG, there have been many classical methods in computational chemistry. Recently, with the development of deep learning, researchers have also proposed some data-driven solutions. In the following, we provide a brief overview of the related work.

\subsection{Classical Methods}

The traditional paradigm of MCG~\cite{hawkins2017conformation} is to first use the conformational search methods to generate a bunch of molecular conformations, and after the energy minimization (a conformation optimization method based on energy) of each conformation, a batch of molecular conformations with low energies are screened out based on the potential energy 
evaluated by density functional theory (DFT)~\cite{parr1980density,rossi2014validation}, semi-empirical DFT~\cite{bannwarth2019gfn2}, or molecular mechanism (MM)~\cite{halgren1996merck,rappe1992uff}.

The core problem of MCG is the conformational search problem. Since there are many rotatable dihedral angles in each molecule, a small change of each rotatable dihedral angle will generate a new molecular conformation. This is a combinatorial explosion problem. It is an urgent problem to generate conformations with lower energy under limited resources.
Several popular conformational search methods have been proposed.
The system search method~\cite{beusen1996systematic, sauton2008ms} will uniformly sample the conformation of each variable dihedral angle at a certain step size, but it is not suitable for molecules with more dihedral angles and molecules containing large rings.
The random search algorithm~\cite{wilson1991applications} randomly perturbs each dihedral angle, and then calculates the acceptance probability based on the Monte Carlo method to decide whether to keep the generated conformation. It has a long convergence time and has repeated visits.
The model-building method~\cite{smellie2003conformational} decomposes large molecules into molecular fragments, extracts the conformation of the molecular fragments from the structural database, and then splices the fragments to form the original molecular conformation. ETKDG\cite{riniker2015better} and OMEGA\cite{hawkins2010conformer} both use this technique. It performs poorly on complex systems and molecules with strong intramolecular interactions. 
The distance geometry method~\cite{spellmeyer1997conformational} regards the molecular conformation as a series of points defined by distance parameters, and uses a geometric search algorithm to search for the best conformation, which is more useful when the distance information is known.
Molecular dynamics simulation ~\cite{tsujishita1997camdas} allows molecules to generate conformations based on Newtonian mechanics under the constraints of molecular force fields, which can generate a large number of molecular conformations, but consumes more computing resources than the above methods.

After the conformational search problem is stated, the generated conformations still need to be evaluated and screened out. The commonly used methods are electronic structure method and force field methods. In terms of energy estimation, force field methods, such as MMFF\cite{halgren1996merck} and UFF\cite{rappe1992uff}, are much less accurate than electronic structure methods. However, quantum mechanical or density functional theory (DFT) methods are time-consuming and unacceptable in the face of large numbers of molecules.

\subsection{Deep Learning methods} 
\label{sec:dl_methods}
Recently, with the development of deep learning, researchers expect to deal with MCG with data-driven solutions, and on the GEOM benchmark, deep learning methods outperformed traditional methods such as EDKTG.
In earlier work, CVGAE~\cite{mansimov2019molecular} used VAE to generate atomic coordinates directly. However, this model could not maintain translation and rotation equivariance of the generated conformations and performed poorly on the benchmark. To ensure such geometrical properties of the conformation, some later works use intermediate structures, such as interatomic distances or torsion angles, to generate conformations. GraphDG\cite{simm2020generative} and CGCF\cite{xu2020learning} use VAE and flow to generate distance matrices, respectively, and then use classical distance geometry techniques to generate conformations iteratively. ConfVAE\cite{xu2021end} design an end-to-end framework to enhance model performance using bilevel optimization. However, minor errors in the predicted distance matrix can cause the generated 3D structures to be unreasonable, such as local structural conflicts that do not fit the triangular inequality. ConfGF\cite{shi2021learning} proposes to learn the gradient of coordinates to alleviate this problem. DGSM\cite{luo2021predicting} takes the non-bonded local and long-range interactions into account when modeling. But they still rely on intermediate structural elements. The method of generating all the atom coordinates in the molecule directly, which avoids the intermediate geometric elements mentioned above, has yet to lose its attention. Similar to the rigid rotor approximation, GeoMol\cite{ganea2021geomol} does not consider unnecessary degrees of freedom and only predicts local 3D structures and torsion angles. DMCG\cite{zhu2022direct} constantly refines the coordinates along model blocks and proposes a loss function that is invariant to the roto-translation of the atom coordinates and the permutation of symmetric atoms. With the advent of diffusion models, methods, such as GeoDiff~\cite{xu2022geodiff} and Torsional Diffusion~\cite{jing2022torsional}, apply this latest technique to the conformation generation task.

\begin{figure}[t]
    \centering
    \includegraphics[width=13.5cm,height=7.5cm]{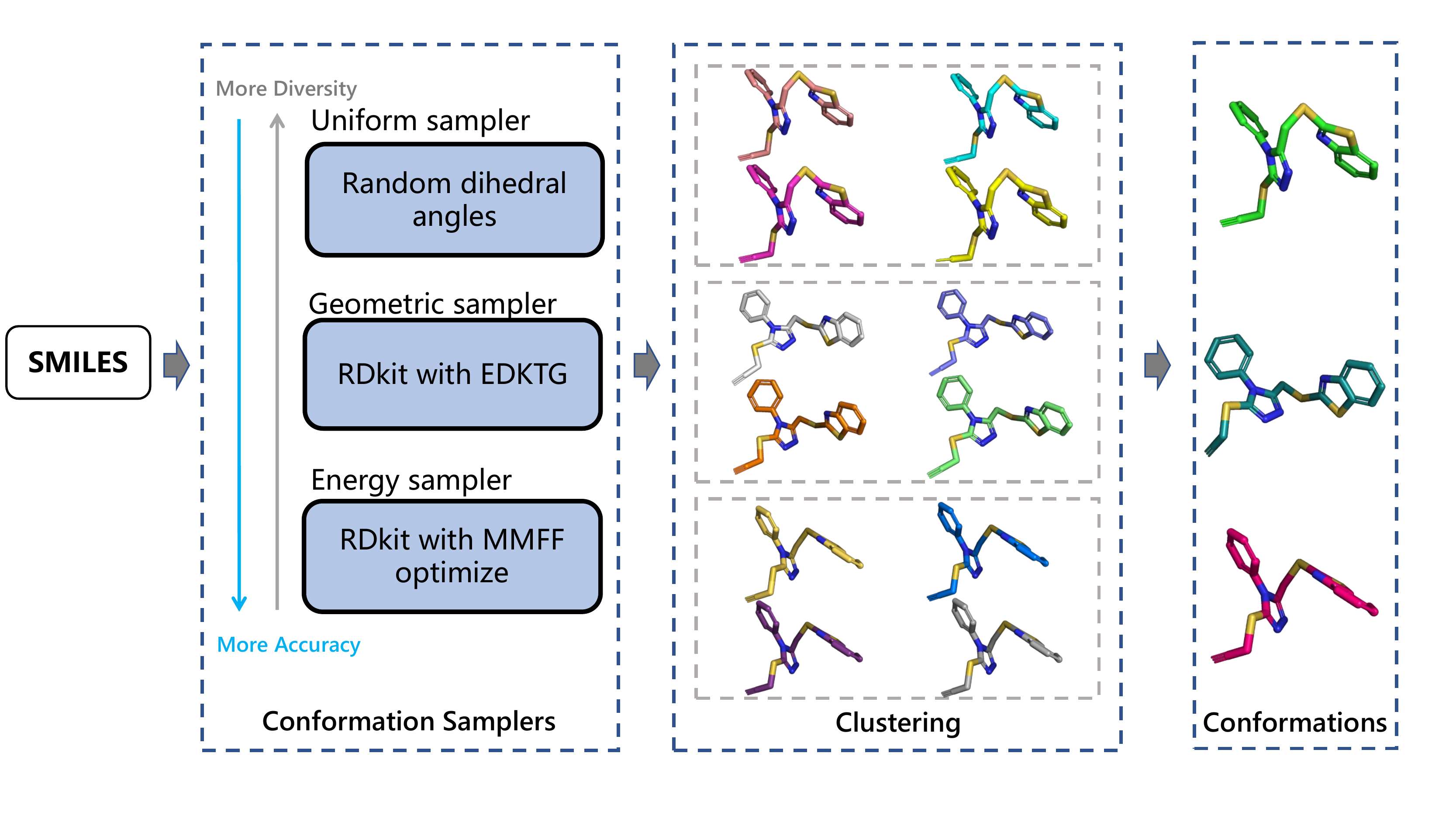}
    \caption{Overview of the proposed RDKit + Clustering algorithm.}
    \label{fig:confg}
\end{figure}

\section{Method}

In this paper, we present a straightforward approach utilizing RDKit\cite{landrum2013rdkit} and clustering post-processing. Our method involves three samplers to produce a range of low-energy conformations that are diverse in structure, as depicted in Fig.~\ref{fig:confg}. The first sampler employs a uniform distribution of dihedral angles in molecules, without taking energy estimation into account. The second sampler, RDKit's default conformation generation setting, uses the EDKTG geometric sampler to assemble conformations with specific molecular fragments based on knowledge and structural databases. Finally, the energy sampler generates more accurate energy estimations for better conformational optimization by utilizing the RDKit and MMFF force field optimization methods.

By integrating those samplers, a simple unsupervised cluster algorithm is applied to select conformations with consideration of diversity and energy. It is noticed that the majority of clusters share conformations with lower energy, and different clusters share conformations with more diversity. 

Specifically, our sampling process involves the uniform, geometric, and energy samplers, which are used in a ratio of 1:1:4, respectively. The number of energy samples, denoted by $N_{e}$, is determined by the formula $N_{e} = min(20N_{ref}, 2000)$, where $N_{ref}$ refers to the number of reference conformations of the current molecule in the test set. The uniform and geometric samplers generate a quarter of the number of samples produced by the energy sampler, with $N_{u}$ and $N_{g}$ representing their respective sample sizes. Once a sufficient number of samples have been generated, we employ the K-means algorithm with $2N_{ref}$ clusters to cluster them, and select the center of each cluster as an output. Prior to clustering, we align and centralize the 3D coordinates of the samples using the kabsch algorithm~\cite{kabsch1993automatic}, followed by flattening the vector of aligned coordinates for K-means. This method provides twice the number of conformations as $N_{ref}$ for metric calculations, consistent with previous research~\cite{xu2020learning, shi2021learning}.

\section{Experiment}

\paragraph{Datasets and setup}
Following the settings in previous works~\cite{xu2020learning, shi2021learning}, we use GEOM-QM9 and GEOM-Drugs~\cite{axelrod2022geom} dataset as the benchmark. The training set is constructed by randomly selecting 40,000 molecules from these two datasets, respectively, and each molecule is taken with 5 possible low-energy conformations sorted by energy. From the remaining data, 200 molecules are randomly selected, and the number of conformations per molecule is limited to between 50 and 500. 22,408 and 14,324 conformations are available in the test set for GEOM-QM9 and GEOM-Drugs, respectively. Since our method (RDKit+Clustering) does not require training, for the inference in the test set, we generate the same number of conformations (twice the number of labeled conformations) as previous works do. 
And we also use the same metrics, Coverage (COV) and Matching (MAT) as in previous works. Higher COV means better diversity, while lower MAT means higher accuracy. Despite the proposal of several new metrics and benchmarks for evaluation in previous works, the absence of a standardized metric, other than COV and MAT, limits our ability to make systematic comparisons on them.
More details about metrics are shown in Appendix~\ref{app:metrics}.

\paragraph{Baselines}
We compare our method with 10 competitive baselines.
RDKit~\cite{riniker2015better}, i.e., original EDKTG with MMFF, is a traditional conformation generation method based on distance geometry. 
The introduction of other baselines can be found in Sec.~\ref{sec:dl_methods}.
A recent model known as torsional diffusion has been introduced \cite{jing2022torsional}, but we have not compared it in our paper due to the following reasons. Firstly, the method employs different splits for training and test data, making an apple-to-apple comparison difficult. Moreover, torsional diffusion is a combination of conventional methods with deep learning, rather than a purely deep learning approach. It relies on RDKit-generated conformations as input, and modifies their torsional angles.

\paragraph{Results}
The results are shown in Table~\ref{confgen}. We report the mean and median of COV and MAT on GEOM-QM9 and GEOM-Drugs datasets. 
ConfVAE~\cite{xu2021end}and DGSM~\cite{luo2021predicting} results are from their papers, respectively. GeoMol\cite{ganea2021geomol} results are from DMCG~\cite{zhu2022direct} paper, which DMCG reproduced after aligning the data split. We found the test sets used in GeoDiff~\cite{xu2022geodiff} and DMCG~\cite{zhu2022direct} are slightly different from baselines (different filtering conditions), so we use the released model parameters of GeoDiff to reproduce the results on the same test set. For DMCG, as model parameters are not released, we use its open-source codes to reproduce from scratch.
Other baseline results are from ConfGF's paper.
As shown in Table~\ref{confgen}, our method exceeds most existing baselines in COV and MAT metrics and achieves comparable performance to SOTA.

\begin{table}[t]

  \caption{Performance on GEOM-QM9 and GEOM-Drugs. The best results are marked \textbf{bold}.}

  \label{confgen}
  \centering
  \small
  \begin{tabular}{l|llll|llll}
    \toprule
    Dataset &  \multicolumn{4}{c}{QM9} &  \multicolumn{4}{c}{Drugs}  \\
     \multirow{2}{*}{Methods} &  \multicolumn{2}{c}{COV($\uparrow$, \%)} & \multicolumn{2}{c}{MAT($\downarrow$, \text{\AA})}&  \multicolumn{2}{c}{COV($\uparrow$, \%)} & \multicolumn{2}{c}{MAT($\downarrow$, \text{\AA})}\\

     & \multicolumn{1}{c}{Mean} & \multicolumn{1}{c}{Median} & \multicolumn{1}{c}{Mean} & \multicolumn{1}{c}{Median} & \multicolumn{1}{c}{Mean} & \multicolumn{1}{c}{Median} & \multicolumn{1}{c}{Mean} & \multicolumn{1}{c}{Median}\\
    \midrule
     RDKit  & 83.26  & 90.78 &  0.3447 & 0.2935 & 60.91 & 65.70  & 1.2026  &  1.1252\\
     CVGAE  & 0.09 &  0.00 &  1.6713 & 1.6088 &  0.00 & 0.00 & 3.0702 & 2.9937\\
     GraphDG & 73.33 & 84.21 & 0.4245  & 0.3973 &  8.27 & 0.00 & 1.9722 &  1.9845 \\
     CGCF & 78.05 & 82.48 & 0.4219 & 0.3900 &  53.96 & 57.06 & 1.2487  & 1.2247 \\
     ConfVAE & 80.42 & 85.31 & 0.4066 & 0.3891 & 53.14 & 53.98 & 1.2392 & 1.2447\\
     ConfGF & 88.49  & 94.13 & 0.2673 & 0.2685 &  62.15  & 70.93 &  1.1629  &  1.1596\\
     GeoMol & 71.26 & 72.00 & 0.3731 & 0.3731 & 67.16 & 71.71 & 1.0875 & 1.0586 \\
     DGSM & 91.49  & 95.92 & 0.2139 & 0.2137 &  78.73  &  94.39  &  1.0154 & 0.9980\\
     GeoDiff & 92.65   & 95.75 & 0.2016 & 0.2006 & 88.45  &  97.09  &  0.8651 & 0.8598 \\
     DMCG & 94.98 & 98.47 & 0.2365 & 0.2312 & \textbf{91.27} & \textbf{100.00} & 0.8287 & 0.7908 \\
    \midrule

    RDKit + Clustering & \textbf{97.65} & \textbf{100.00} & \textbf{0.1902} & \textbf{0.1818} & 87.93 & \textbf{100.00} & \textbf{0.8086} & \textbf{0.7838} \\
    
    \bottomrule
  \end{tabular}
\end{table}

\paragraph{Ablation study}
As mentioned in Sec.~\ref{intro}, we hypothesize that more diverse conformations could easily achieve better results in GEOM. To demonstrate this explicitly, we design ablation studies for the sample size and sampler types. For the sample size, we set the energy sampler's sample size to twice, five times, and ten times the number of reference conformations $N_{ref}$. The sampling ratio of 3 samplers remains the same as before. For the sampler type, we remove one of the 3 samplers separately in the experiment. The sampling ratio also remains the same. 

The results are shown in Table ~\ref{abla}. From the results, the smaller the number of samples, the worse the performance, although there is some marginal utility. In terms of samplers, the energy sampler mainly improves the accuracy of the generated conformations while decreasing the diversity; the uniform sampler mainly improves the diversity while making the accuracy decrease; geometric samplers have a positive impact on both accuracy and diversity.

\begin{table}[h]

  \caption{Ablation studies for the number of samples and sampler type on GEOM-QM9. The best results are marked \textbf{bold}.}

  \label{abla}
  \centering
  \small
  \begin{tabular}{l|llll}
    \toprule
    Dataset &  \multicolumn{4}{c}{QM9}   \\
     \multirow{2}{*}{Methods} &  \multicolumn{2}{c}{COV($\uparrow$, \%)} & \multicolumn{2}{c}{MAT($\downarrow$, \text{\AA})}\\
     & \multicolumn{1}{c}{Mean} & \multicolumn{1}{c}{Median} & \multicolumn{1}{c}{Mean} & \multicolumn{1}{c}{Median}\\
    \midrule
     2$N_{ref}$  & 91.23  & 96.87 &  0.2223 & 0.2171 \\
     4$N_{ref}$  & 95.19  & 99.58 &  0.1988 & 0.1901 \\
     10$N_{ref}$  & 96.49  & 100.00 &  0.1941 & 0.1857 \\
     w/o Uniform sampler  & 93.56  & 99.55 &  \textbf{0.1867} & \textbf{0.1643} \\
     w/o Geometric sampler  & 97.42  & 100.00 &  0.1984 & 0.1930 \\
     w/o Energy sampler  & \textbf{98.01}  & 100.00 &  0.2511 & 0.2434 \\
    \midrule
    RDKit + Clustering & 97.65 & \textbf{100.00} & 0.1902 & 0.1818 \\
    
    \bottomrule
  \end{tabular}
\end{table}

\paragraph{Discussion}
While our approach surpasses most of the baselines, we have doubts that the benchmark utilized can fulfill the needs of real-world conformation generation tasks. The ablation studies indicate that this benchmark can be improved by generating more diverse conformations. Furthermore, the requirements for conformations differ depending on the applications. For instance, in molecular docking, we require conformations that are similar to the binding conformation. The ensemble of bioactive conformations must be taken into account when using the conformation generation model in drug design, which is overlooked by GEOM. In contrast, for quantum property prediction, we require conformations that are close to the DFT-optimized conformations. The accuracy of GEOM's semi-empirical DFT is insufficient to meet these requirements. Therefore, we need to benchmark the impact of conformation generation according to the downstream applications' needs or their final performance. 
\section{Conclusion}

Our study presents a simple algorithm that utilizes traditional approaches and has demonstrated superior performance compared to most deep learning models. As a result, we conducted an investigation into the current settings utilized in deep learning methods MCG, and found that the current benchmark for MCG may contain inaccuracies. We recommend that future research should focus on the end applications in various MCG-related downstream applications, given that benchmarking the performance of MCG alone is inherently challenging. We are confident that with a reasonable benchmark, deep learning-based methods can significantly contribute towards the development of effective MCG models for specific tasks such as bioactive conformation generation, general conformation optimization, and conformation scoring.

\subsubsection*{Acknowledgments}
We thank Yuejiang Yu, Shuqi Lu, Junhan Chang, Xi Chen and many colleagues in DP Technology for their great help in this project.

{
\small
\printbibliography
}

\clearpage
\appendix
\section{Metrics in GEOM Dataset}
\label{app:metrics}
\subsection{RMSD}
Root of Mean Squared Deviations (RMSD) is widely used to evaluate the difference between conformations in molecules. Before computing RMSD, the generated conformation is first aligned with the reference one, and the function $\Phi$ aligns conformations by applying rotations and translations to them:
\begin{equation}
\text{RMSD}(\boldsymbol{R},\hat{\boldsymbol{R}})= \min_\Phi(\frac{1}{n}\sum_{i=1}^n||\Phi(\boldsymbol{R}_i)-\hat{\boldsymbol{R}}_i||^2)^{\frac{1}{2}}
\end{equation}
where $\boldsymbol{R}$ and $\hat{\boldsymbol{R}}$ are the generated and reference conformation, $i$ is the $i$-th heavy atom, and $n$ is the number of heavy atoms.
\subsection{COV \& MAT}
Coverage (COV) and Matching (MAT) is used to evaluate the performance of the conformation generation model in GEOM. Higher COV means better diversity, while lower MAT means higher accuracy. Formally, COV and MAT are denoted as:
\begin{equation}
\label{eq:cov}
\text{COV}(S_{g},S_{r})= \frac{\left|\left\{ \boldsymbol{R} \in S_{r} | \text{RMSD}(\boldsymbol{R},\hat{\boldsymbol{R}})< \delta,\hat{\boldsymbol{R}} \in S_{g}\right\}\right|}{|S_{r}|}
\end{equation}

\begin{equation}
\label{eq:mat}
\text{MAT}(S_{g},S_{r})= \frac{1}{|S_{r}|} \sum_{\boldsymbol{R} \in S_{r}} \min_{\hat{\boldsymbol{R}}\in S_{g}} \text{RMSD}(\boldsymbol{R},\hat{\boldsymbol{R}})
\end{equation}
where $S_{g}$ and $S_{r}$ are the set of generated and reference conformations, respectively, and $\delta$ is a given RMSD threshold. Following previous work~\cite{xu2020learning,shi2021learning}, for GEOM-QM9, the threshold is 0.5$\text{\AA}$, and for GEOM-Drugs, the threshold value is 1.25$\text{\AA}$.
\end{document}